# Physical Layer Security Enhancement for Satellite Communication among Similar Channels: Relay Selection and Power Allocation

Shuai Han, Xiangxue Tai
Harbin Institute of Technology, China

**Abstract:**

Channels of satellite communication are usually modeled as Rician fading channels with very large Rician factor or Gaussian channels. Therefore, when a legitimate user is close to an eavesdropping user, the legitimate channel is approximately the same as the eavesdropping channel. The physical layer security technology of traditional terrestrial wireless communication mainly takes advantage of the difference between the legitimate channel and the eavesdropping channel; thus, it is not suitable for satellite communication. To implement secure communication in similar channels for satellite communications, a secure communication model based on collaboration of the interference relay of the satellite physical layer is proposed. Relay selection and power allocation are further studied to enhance the security performance of the satellite communication system based on the model. The relay selection standard under known instantaneous channel state information (CSI) and statistical CSI conditions is theoretically derived, thereby accomplishing minimization of the probability of secrecy relay. In addition, the power allocation factor is optimized based on minimization of the secrecy outage probability. Moreover, a power allocation method based on the statistical CSI is presented. The secrecy outage probability performance of each relay selection criterion and power allocation scheme are analyzed via a simulation.

## 1 Introduction

The security problems that accompany the wide-area openness of satellite communication are far more serious than those of terrestrial networks. With the development of quantum computing, the encryption of traditional information layers faces the risk of being brute-force cracked. Nevertheless, the theory of physical layer security will become a powerful means of ensuring the security of satellite communications.

To date, there has been extensive research regarding physical layer security technology. Beamforming is a basic technique for implementing secure communication of the physical layer of the system and is applied extensively to achieve security in specific communication scenarios. Reference [1] analyzed the physical layer security of a multi-input-multi-output-transmitting beamforming system in the eavesdropping channel based on a maximum-ratio combining (MRC) receiver. Reference [2] presented a novel frequency diverse array (FDA) beamforming method. The highly correlated channels of legitimate and eavesdropping users are decorrelated via the frequency offset of the array antenna. Reference [3] proposed a new, location-based beamforming method with a focus on eavesdropping channels. Like the beamforming technique, the artificial noise technique can also effectively enhance the secrecy performance of the system. The principal idea of the artificial noise technique is to incorporate appropriate pseudo-random noise into the transmitted signal to interfere with the eavesdropper, such that the signal-receiving quality of the eavesdropper is reduced, thereby enhancing the secrecy capability. With multiple available antennas from the transmitter, the information and artificial noise (AN) can be transmitted simultaneously [4][5]. S. Goel and R. Negi pioneered using AN methods to enhance the channel capacity [6]. Reference [7] proposed a new strategy based on ordinal eigenvalues of the Wishart matrix. The message is encoded in the *s* strongest characteristic subchannels, and the AN signal is generated in the remaining space. The complex non-central Wishart distribution is applied to extend this strategy to the Rician channel. The performance of a multiple-input and multiple-output (MIMO) communication system is simulated in a Rician fading environment to study the influence of the Rician factor on the security capacity. Reference [8] presents a transmission method that maximizes the secrecy outage probability in the case of multi-antenna transmission. The accuracy of the channel state information (CSI) corresponding to the eavesdropping channel affects the application of physical layer security techniques. Different AN design methods are employed depending on whether the eavesdropping channel state is known. The design of directional isotropic AN is utilized in the case in which the CSI of the eavesdropper is unknown [9] [10]. In addition, the AN technique is often combined with beamforming to achieve physical layer security. In reference [11], an AN optimization method that assists and coordinates multiple cells in beamforming is studied for multi-user systems. All base stations of beamforming and AN vector are collectively optimized to reduce the total transmitting power to the largest extent possible while simultaneously guaranteeing the quality of service (QoS) for authorized users and avoiding interception of information by unauthorized users. The optimal AN-assisted beamforming method is also studied in reference [12]. Several existing results regarding single-antenna eavesdroppers are extended to multi-antenna eavesdroppers; the results proved that the traditional zero-space AN plan is indeed the best for any parameter of the system. Reference [13] proposed an optimized secure multi-antenna transmission method based on AN-assisted secure beamforming for the situation in which the feedback from a single-antenna receiver is limited. Reference [14] designed an AN-assisted transmission strategy for a multiple-input single-output multi-antenna eavesdropper (MISOME). The transmission and AN covariance are collectively optimized owing to the maximization of the secure rate. The co-frequency co-time full-duplex technology can be applied to transmit AN for the purpose of interfering with the eavesdropping node and thereby enhance the secrecy performance of the system. The full-duplex technology considers two primary scenarios for the physical layer security. (1) The legitimate receiving end has full duplex functionality; it receives the signals coming from the transmitting end while transmitting the AN to interfere with the eavesdropper. (2) In the wiretap model with relays, it deploys the full-duplex technology on the relay node. The relay node receives the signal while transmitting the interference signal to interfere the activity of the eavesdropper [15] [16].

The physical layer security technique for terrestrial wireless communication stated above utilizes the difference between legitimate and eavesdropping channels to achieve secure communication. However, in satellite communications, because the altitude of satellite is much greater than that of a legitimate receiver, almost no blocking obstacles or reflection are present in the transmission. The power of the direct signal is much greater than that of a multi-path signal. Consequently, the channel is essentially related to the transmission loss, which is determined by the distance. When a legitimate user is close to an eavesdropping user, the distance between the legitimate and eavesdropping users is much less than the distances from the satellite to both legitimate and eavesdropping users, and it can almost be ignored. Hence, the CSI of the legitimate channel is approximately equal to the CSI of the eavesdropping channel. Therefore, physical layer techniques for traditional terrestrial wireless communication cannot be directly applied to the security of satellite communication.

This paper presents research regarding the physical layer security in the case in which legitimate and eavesdropping channels are approximately equal in satellite communication. The co-frequency co-time cooperative interference relay is incorporated to transmit AN interference signals that only affect the eavesdropping user by reducing the quality of the signal received while ensuring its quality for the legitimate user, thereby ensuring the secrecy capability of the system and achieving the security of the communication. To further improve the secrecy performance of the system, physical layer security enhancement techniques, including relay selection and power allocation, are studied, with a focus on the circumstances detailed above. The relay selection mainly concerns optimal selection among the multiple relays that can be accessed by the satellite communication system such that the secrecy outage probability performance of the system is optimized. Power allocation mainly concerns the case in which the total transmitting power of the satellite and cooperative interference relays is fixed. By making the most effective use of power allocation between the satellite and relay, the performance in terms of the secrecy outage probability is optimized.

Notations:

## 2 System models

To resolve the physical layer security problem in the case of similar satellite communication channels, incorporation of a near-ground co-frequency co-time cooperative interference relay is considered. The near-ground co-frequency co-time cooperative interference relay transmits an interference signal that theoretically has no effect on the legitimate receiving end according to certain strategy. This approach significantly reduces the quality of the signal received by an eavesdropping receiver, while the signal quality remains substantially unchanged for legitimate users, thereby ensuring the secrecy capability of the system and achieving the security of the system.

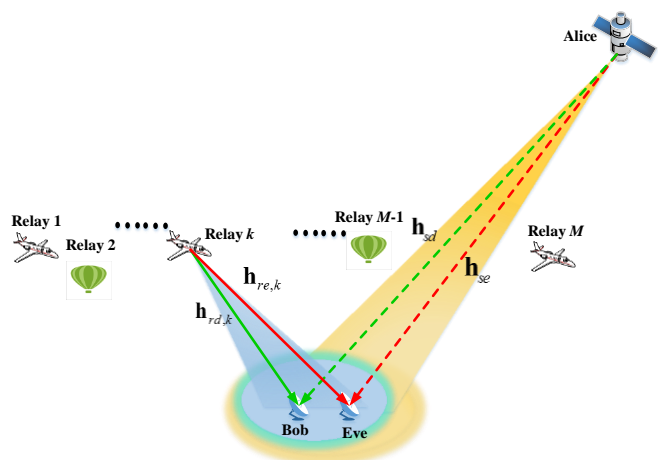

Figure 1 Satellite-ground physical layer secure communication system model

Figure 1 shows the satellite-ground physical layer secure communication system model. The model contains a satellite (Alice), a legitimate receiver end (Bob), an eavesdropping receiver end (Eve), and M near-ground co-frequency co-time cooperative interference relays $\{\text{Relay } k\}_k^M$ with respect to the satellite. Assume that Bob is close to Eve. In actual satellite communications, the satellite-ground channel is a Rician fading channel with very large Rician factor. The relay is near the ground; thus, the channel relayed to the ground receiving end is modeled as a Rician and Rayleigh channel. Therefore, when Bob is near Eve, the channels from Alice to Bob and from Alice to Eve are approximately equal, whereas the channels relaying to Bob and Eve differ. When data are must be transmitted between the satellite and Bob, a selection from the M accessible relays near Bob must be made according to a relay selection strategy, and the selected relay participates in the data transmission of the satellite communication system. Alice transmits a confidential data signal; the relay transmits an AN interference signal that is orthogonal to the channel between the relay and the legitimate receiving end. Both Bob and Eve can receive cooperative interference relays from Alice and the relay. Assume that the numbers of transmitting antennas for the satellite and relay are $N_s$ and $N_r$, respectively, the receiving antennas of Bob and Eve are single antennas, and the satellite's and relay's transmitting powers are $P_s$ and $P_r$, respectively. Then, in the case in which the relay selected by the satellite communication system that participates in data transmission is relay $k$, assuming the legitimate and eavesdropping receiving ends have the same gain, the signals received by Bob and Eve are

$$y_{d,k} = \sqrt{P_s}\mathbf{h}_{sd}^H\mathbf{x}_s + \sqrt{P_r}\mathbf{h}_{rd,k}^H\mathbf{x}_{z,k} + n_d \quad (1)$$
$$y_{e,k} = \sqrt{P_s}\mathbf{h}_{se}^H\mathbf{x}_s + \sqrt{P_r}\mathbf{h}_{re,k}^H\mathbf{x}_{z,k} + n_e \quad (2)$$

respectively, in which $y_{d,k}$ represents the signal received by Bob when the selected relay is relay k; $y_{e,k}$ indicates the signal received by Eve when the selected relay is relay k; $\mathbf{h}_{sd}$ is the CSI vector from Alice to Bob, $\mathbf{h}_{sd} \in \mathbb{C}^{N_s \times 1}$; $\mathbf{h}_{se}$ is the CSI vector from Alice to Eve, $\mathbf{h}_{se} \in \mathbb{C}^{N_s \times 1}$; $\mathbf{h}_{rd,k}$ is the CSI vector between the relay and Bob, $\mathbf{h}_{rd,k} \in \mathbb{C}^{N_r \times 1}$; and $\mathbf{h}_{re,k}$ is the CSI vector from the relay to Eve, $\mathbf{h}_{re,k} \in \mathbb{C}^{N_r \times 1}$. $n_d$ is the complex additive Gaussian random noise at Bob's end, $n_d \in CN(0, \sigma_d^2)$, and $n_e$ is the complex additive Gaussian random noise of the eavesdropping node at Eve's end, $n_e \in CN(0, \sigma_e^2)$. $\mathbf{x}_s$ represents the confidential data signal transmitted by Alice, $\mathbf{x}_s \in \mathbb{C}^{N_s \times 1}$, for which the power is 1. $\mathbf{x}_{z,k}$ indicates the AN interference signal transmitted to Eve by the cooperative interference relay $k$, which is the pseudo-random complex Gaussian noise orthogonal to the channel between relay $k$ and Bob; that is, $\mathbf{x}_{z,k}$ is located in the null space of channel $\mathbf{h}_{rd,k}$, $\mathbf{x}_{z,k} \in \mathbb{C}^{N_r \times 1}$, for which the power is equal to 1. $\mathbf{x}_{z,k}$ can be specifically expressed as

$$\mathbf{x}_{z,k} = \mathbf{G}_{\mathbf{h}_{rd,k}}\mathbf{z}_k \quad (3)$$

in which $\mathbf{G}_{\mathbf{h}_{rd,k}}$ is the normalized orthogonal basis matrix of the null space for channel $\mathbf{h}_{rd,k}$, $\mathbf{G}_{\mathbf{h}_{rd,k}} \in \mathbb{C}^{N_r \times N_r - 1}$, that is, $\mathbf{h}_{rd,k}^H \mathbf{G}_{\mathbf{h}_{rd,k}} = 0_{1 \times N_r - 1}$. $\mathbf{z}_k$ is the pseudo-random complex Gaussian noise, $\mathbf{z}_k \in \mathbb{C}^{N_r - 1}$.

According to Equations (1), (2), and (3), the ratios of signal to interference at Bob and Eve are $\gamma_{d,k}$ and $\gamma_{e,k}$, respectively:

$$\gamma_{d,k} = \frac{P_s \|\mathbf{h}_{sd}\|^2}{\sigma_d^2} \quad (4)$$

$$\gamma_{e,k} = \frac{P_s \|\mathbf{h}_{se}\|^2}{P_r \|\mathbf{h}_{re.k}^H \mathbf{G}_{\mathbf{h}_{rd.k}}\|^2 + \sigma_e^2} \quad (5)$$

Then, when the relay is relay k, the instantaneous secrecy capacity at Bob is

$$C_{s,k} = [\log_2(1+\gamma_{d,k}) - \log_2(1+\gamma_{e,k})]^+ \quad (6)$$

In the case in which the Gaussian noise power is equal to $\sigma_d^2 = \sigma_e^2 = \sigma^2$ at the locations of Bob and Eve, because $\|\mathbf{h}_{sd}\|^2 = \|\mathbf{h}_{se}\|^2$, $P_r \|\mathbf{h}_{re}^H \mathbf{G}_{\mathbf{h}_{rd}}\|^2 > 0$, so $\gamma_d > \gamma_e$; therefore, $C_s > 0$. This result indicates that the model can achieve secure communication. Additionally, in the case of a high signal-noise ratio, as $P_s, P_r \to \infty$, $\gamma_d \to \infty$ and $\gamma_d \to \rho \frac{\|\mathbf{h}_{se}\|^2}{\|\mathbf{h}_{re}^H \mathbf{G}_{\mathbf{h}_{rd}}\|^2}$ ($\rho = P_s/P_r$ is a finite value); thus, $C_s \to \infty$. This result shows that the model can attain an instantaneous secrecy capacity that is arbitrarily large when the transmitting power from Alice is sufficiently large.

### 3 Relay selections

When multiple relay nodes are present in the system, Equation (6) reveals that the selection of the relay node to participate in the data transmission will affect the secrecy performance of the system; thus, relay selection is an important technique for the enhancement of system security. Therefore, this section will study how the relay selection affects the enhancement in the secrecy performance of the system.

#### A. Relay selection based on instantaneous CSI

The channel of satellite communication is a slowly fading channel. The secrecy outage probability (SOP) is an important indicator for the measurement of the secrecy performance of slowly fading channels. The lower the SOP is, the better the secrecy performance of the system. Therefore, optimization of the relay selection based on minimization of the SOP is meaningful.

The SOP is the probability that the instantaneous secrecy capacity is less than a certain threshold. The secret communication should be interrupted once the instantaneous secrecy capacity is below this threshold; otherwise, the communication will not be safe. Assume that the minimum rate of secure system transmission is $R_s$ when the relay is relay $k$; then, the SOP at Bob's end can be expressed as

$$P_{out,k} = Pr[C_{s,k} < R_s] \quad (7)$$

where $Pr[\bullet]$ represents solving for the probability.

The selected relay should be the relay that minimizes the SOP of the system among the $M$ relays. Therefore, the optimal relay $k^*$ can be expressed as

$$k^* = \underset{k=1,2,\ldots,M}{\arg\min} P_{out,k} \quad (8)$$

Substituting Equation (7) into Equation (8) yields

$$k^* = \underset{k=1,2,\ldots,M}{\arg\max} C_{s,k} \quad (9)$$

Equation (9) reveals that the optimal relay $k^*$ obtained based on minimization of the SOP is equivalent to the optimal relay $k^*$ determined based on maximization of the instantaneous secrecy capacity.

Substituting Equations (4), (5) and (6) into Equation (9), the optimal relay $k^*$ can be expressed as

$$k^* = \underset{k=1,2,\ldots,M}{\arg\max}[\log_2(1+\frac{P_s \|\mathbf{h}_{sd}\|^2}{\sigma_d^2}) \\ -\log_2(1+\frac{P_s \|\mathbf{h}_{se}\|^2}{P_r \|\mathbf{h}_{re.k}^H \mathbf{G}_{\mathbf{h}_{rd.k}}\|^2 + \sigma_e^2})] \quad (10)$$

Since $P_s$, $P_r$, $\|\mathbf{h}_{sd}\|^2$, $\|\mathbf{h}_{se}\|^2$, $\sigma_d^2$ and $\sigma_e^2$ are not related to relay $k$, Equation (10) is equivalent to

$$k^* = \arg\max_{k=1,2,\ldots,M} \|\mathbf{h}_{re.k}^H \mathbf{G}_{\mathbf{h}_{rd.k}}\|^2 \quad (11)$$

According to Equation (11), it is evident that the optimal relay $k^*$ is determined by the instantaneous CSI of channels $\{\mathbf{h}_{re.k}\}_1^M$ and $\{\mathbf{h}_{rd.k}\}_1^M$. The optimal relay $k^*$ must be continuously updated as the instantaneous CSI of channels $\{\mathbf{h}_{re.k}\}_1^M$ and $\{\mathbf{h}_{rd.k}\}_1^M$ changes.

**B. Relay selection based on the statistical CSI**

In part A, the optimal relay $k^*$ is determined based on the instantaneous CSI of the channels $\{\mathbf{h}_{re.k}\}_1^M$ and $\{\mathbf{h}_{re.k}\}_1^M$ from $\{\text{Relay } k\}_1^M$ to Bob and Eve. However, errors are present in channel estimation, and it is difficult for satellite communication systems to obtain accurate CSI for $\{\mathbf{h}_{re.k}\}_1^M$ and $\{\mathbf{h}_{re.k}\}_1^M$; thus, satellite communication systems may not have the ability to attain instantaneous CSI. In addition, the instantaneous CSI for $\{\mathbf{h}_{re.k}\}_1^M$ and $\{\mathbf{h}_{re.k}\}_1^M$ changes continuously; therefore, relay selection must be performed continuously by the satellite communication system. This will increase the signaling overhead, and moreover, performing relay-handover access continuously will increase the delay of data transmission. To remedy the drawbacks of the relay selection method based on known instantaneous CSI presented in Part A while improving the actual realizability, the relay selection should be based on statistical CSI.

According to the optimal statistical CSI-based relay selection standard, when the instantaneous CSI cannot be attained, the optimization function can be modified to

$$\begin{aligned} k^* &= \underset{k=1,2,\ldots,M}{\arg\max} Pr\left[\|\mathbf{h}_{re.k}^H \mathbf{G}_{\mathbf{h}_{rd.k}}\|^2 > r\right] \\ &= \underset{k=1,2,\ldots,M}{\arg\max} 1 - F_{\|\mathbf{h}_{re.k}^H \mathbf{G}_{\mathbf{h}_{rd.k}}\|^2}(r) \quad (12) \\ &= \underset{k=1,2,\ldots,M}{\arg\min} F_{\|\mathbf{h}_{re.k}^H \mathbf{G}_{\mathbf{h}_{rd.k}}\|^2}(r) \end{aligned}$$

The meaning of Equation (12) is to select the relay with the highest probability whose random variable $\|\mathbf{h}_{re.k}^H \mathbf{G}_{\mathbf{h}_{rd.k}}\|^2$ ($k=1,2, \ldots, M$) is greater than any other relays with the same threshold values.

Assume that the channels of relay $\{\text{Relay } k\}_1^M$ to eavesdropping and legitimate receivers are independent Rayleigh channels with the same distribution. Then, random variables $\|\mathbf{h}_{re.k}\|^2$ and $\|\mathbf{h}_{rd.k}\|^2$ both obey $\chi^2$ distributions that have the same number of degrees of freedom, $2N_r$. Additionally, since $\mathbf{G}_{\mathbf{h}_{rd.k}}$ is the normalized orthogonal basis matrix of the null space for channel $\mathbf{h}_{rd,k}$, $\mathbf{G}_{\mathbf{h}_{rd.k}} \in \mathbb{C}^{N_r \times N_r - 1}$, the random variable $\|\mathbf{h}_{re.k}^H \mathbf{G}_{\mathbf{h}_{rd.k}}\|^2$ obeys the $\chi^2$ distribution that has $2N_r - 2$ degrees of freedom. Because the number of degrees of freedom, $2N_r - 2$, is an even number, the cumulative distribution function (CDF) of the random variable $\|\mathbf{h}_{re.k}^H \mathbf{G}_{\mathbf{h}_{rd.k}}\|^2$ is

$$F_{\|\mathbf{h}_{re.k}^H \mathbf{G}_{\mathbf{h}_{rd.k}}\|^2}(r) = \begin{cases} 1 - e^{-\frac{r}{2\sigma_{re.k}^2}} \sum_{m=0}^{N_r-2} \frac{1}{m!}\left(\frac{r}{2\sigma_{re.k}^2}\right)^m, & r > 0 \\ 0, & r \leq 0 \end{cases}$$
(13)

where $\sigma_{re.k}^2 = \dfrac{E[\|\mathbf{h}_{re.k}\|^2]}{2N_r}$.

Differentiating the CDF of $\|\mathbf{h}_{re.k}^H \mathbf{G}_{\mathbf{h}_{rd.k}}\|^2$

with respect to $\sigma_{re,k}^2$ gives the first derivative of $F_{\|\mathbf{h}_{re,k}^H \mathbf{G}_{\mathbf{h}_{rd,k}}\|^2}(r)$ with respect to $\sigma_{re,k}^2$, which can be expressed as

$$\frac{dF_{\|\mathbf{h}_{re,k}^H \mathbf{G}_{\mathbf{h}_{rd,k}}\|^2}(r)}{d\sigma_{re,k}^2} = -e^{-\frac{r}{2\sigma_k^2}} \frac{1}{\sigma_{re,k}^2} \frac{1}{(N_r-2)!} \left(\frac{r}{2\sigma_{re,k}^2}\right)^{N_r-1}, \quad r>0 \quad (14)$$

Equation (14) reveals that the first derivative of $F_{\|\mathbf{h}_{re,k}^H \mathbf{G}_{\mathbf{h}_{rd,k}}\|^2}(r)$ with respect to $\sigma_{re,k}^2$, $\frac{dF_{\|\mathbf{h}_{re,k}^H \mathbf{G}_{\mathbf{h}_{rd,k}}\|^2}(r)}{d\sigma_k^2}$, is always less than 0. Thus, $F_{\|\mathbf{h}_{re,k}^H \mathbf{G}_{\mathbf{h}_{rd,k}}\|^2}(r)$ is a monotonically decreasing function with respect to $\sigma_{re,k}^2$; that is, the greater the value of $\sigma_{re,k}^2$, the smaller the value of $F_{\|\mathbf{h}_{re,k}^H \mathbf{G}_{\mathbf{h}_{rd,k}}\|^2}(r)$. Therefore, Equation (12) can be converted into

$$k^* = \underset{k=1,2,\ldots,M}{\arg\max} \, \sigma_{re,k}^2 \quad (15)$$

Substituting $\sigma_{re,k}^2 = \frac{E[\|\mathbf{h}_{re,k}\|^2]}{2N_r}$ into Equation (15) gives the statistical CSI-based relay selection criterion as

$$k^* = \underset{k=1,2,\ldots,M}{\arg\max} \, E\left[\|\mathbf{h}_{re,k}\|^2\right] \quad (16)$$

Using Equation (16), the relay that has the maximum statistical CSI for the eavesdropper channel can be selected.

## 4 Power allocation

After determining the relay to be involved in the secure data transmission, the entire secure transmission path of the satellite communication system includes a satellite, a co-frequency co-time cooperative interference relay, an eavesdropper and a legitimate receiver. Then, the transmitting powers of the satellite and relay contribute to the power overhead of the satellite communication system. Therefore, to minimize the energy consumption of the entire system, it is necessary to constrain the total transmitting power of the satellite and relay. When the total transmitting power of the satellite and relay is determined, the transmitting power allocation between the satellite and relay may affect the secrecy performance of the system. For this reason, this section will further study the transmitting power allocation between satellites and relays to enhance the secrecy performance of the system.

Assume that the selected secure transmission participating relay is relay $k$, the total transmitting power constraint of the satellite and relay $k$ is $P$, the transmitting power $P_s$ of Alice is $\alpha P$, and the transmitting power $P_r$ of relay $k$ is $(1-\alpha)P$; $\alpha$ is defined by the power allocation factor, which represents the ratio of the transmitting power from Alice to the total transmitting power of Alice and the relay, $\alpha \in (0,1)$. Then, substituting $P_s = \alpha P$ and $P_r = (1-\alpha)P$ into Equation (6) gives the instantaneous secrecy capacity of the system, which is expressed as

$$C_{s,k}(\alpha) = \log_2(1 + \frac{\alpha P \|\mathbf{h}_{sd}\|^2}{\sigma_d^2}) \\ - \log_2(1 + \frac{\alpha P \|\mathbf{h}_{se}\|^2}{(1-\alpha)P \|\mathbf{h}_{re,k}^H \mathbf{G}_{\mathbf{h}_{rd,k}}\|^2 + \sigma_e^2}) \quad (17)$$

### A. Uniform power allocation

The simplest power distribution is uniform power allocation. The physical layer security model of the satellite communication system, uniform power allocation between Alice and relay $k$, is studied. This means that the transmitting powers of Alice and relay $k$ are equal; the power allocation factor is then $\alpha=0.5$.

The advantages of uniform power allocation between Alice and relay $k$ are no relevant channel information being needed for power allocation, less requirements for satellite communication systems, and reduced signaling overhead for Alice and Relay *k*. However, when applying uniform power allocation, the power allocation factor does not depend on CSI about the satellite communication system. Therefore, optimal results in terms of the secrecy performance of the system cannot be achieved. For the satellite communication system to attain better secrecy performance, it is necessary to optimize the power allocation between Alice and relay *k*.

B. Optimal power allocation

The optimization of power allocation in this section is still based on minimization of the SOP. Therefore, the optimal power allocation factor α* should minimize the SOP. The optimization problem can be expressed as

$$\min P_{out,k}(\alpha) \tag{18}$$
$$\text{s.t. } 0 < \alpha < 1$$

where $P_{out,k}(\alpha) = Pr[C_{s,k}(\alpha) < R_s]$. Following the same logic as in section 3, maximization of instantaneous secrecy capacity yields the minimum SOP. Thus, the optimal power allocation factor α* is expressed as

$$\alpha^* = \arg\max_{0<\alpha<1} C_{s,k}(\alpha) \tag{19}$$

Let $\gamma_{sd} = \dfrac{P\|\mathbf{h}_{sd}\|^2}{\sigma_d^2}$, $\gamma_{rd} = \dfrac{P\|\mathbf{h}_{rd,k}\|^2}{\sigma_d^2}$,

$\gamma_{se} = \dfrac{P\|\mathbf{h}_{se}\|^2}{\sigma_e^2}$, and $\gamma_{re} = \dfrac{P\|\mathbf{h}_{re.k}^H \mathbf{G}_{\mathbf{h}_{rd.k}}\|^2}{\sigma_e^2}$; substituting Equation (17) into Equation (19), the optimal power allocation factor can be rewritten as

$$\alpha^* = \arg\max_{0<\alpha<1} \log_2\left(\dfrac{1+\alpha\gamma_{sd}}{1+\dfrac{\alpha\gamma_{se}}{(1-\alpha)\gamma_{re}+1}}\right)$$
$$= \arg\max_{0<\alpha<1} \dfrac{1+\alpha\gamma_{sd}}{1+\dfrac{\alpha\gamma_{se}}{(1-\alpha)\gamma_{re}+1}} \tag{20}$$

Since $\|\mathbf{h}_{sd}\|^2 = \|\mathbf{h}_{se}\|^2$ and $\sigma_d^2 = \sigma_e^2$, $\gamma_{sd} = \gamma_{se}$. Therefore, Equation (20) is equivalent to

$$\alpha^* = \arg\max_{0<\alpha<1} F(\alpha) \tag{21}$$

where

$$F(\alpha) = \dfrac{-\gamma_{sd}\gamma_{re}\alpha^2 + (\gamma_{sd}\gamma_{re}+\gamma_{sd}-\gamma_{re})\alpha+\gamma_{re}+1}{(\gamma_{sd}-\gamma_{re})\alpha+\gamma_{re}+1}.$$

According to Equation (21), solving for the optimal power allocation factor α* corresponds to determining the maximum value of function $F(\alpha)$. Thus, the maximization and minimization of $F(\alpha)$ must be analyzed. Therefore, a differentiation analysis of $F(\alpha)$ is conducted. The first derivative of $F(\alpha)$ can be expressed as

$$\dfrac{\mathrm{d}F(\alpha)}{\mathrm{d}\alpha}$$
$$= \dfrac{[-\gamma_{sd}\gamma_{re}(\gamma_{sd}-\gamma_{re})]\alpha^2 - 2\gamma_{sd}\gamma_{re}(\gamma_{re}+1)\alpha+\gamma_{sd}\gamma_{re}(\gamma_{re}+1)}{[\alpha\gamma_{sd}+(1-\alpha)\gamma_{re}+1]^2}$$
(22)

The second derivative of $F(\alpha)$ can be expressed as

$$\dfrac{\mathrm{d}^2 F(\alpha)}{\mathrm{d}\alpha^2} = \dfrac{-2\gamma_{sd}\gamma_{re}(\gamma_{re}+1)(\gamma_{sd}+1)}{[\alpha\gamma_{sd}+(1-\alpha)\gamma_{re}+1]^3} \tag{23}$$

Since $\gamma_{sd} > 0$ and $\gamma_{re} > 0$, $-2\gamma_{sd}\gamma_{re}(\gamma_{re}+1)(\gamma_{sd}+1) < 0$. At the same time, because $0 < \alpha < 1$, $\alpha\gamma_{sd} > 0$ and $(1-\alpha)\gamma_{re}$. Hence, $[\alpha\gamma_{sd}+(1-\alpha)\gamma_{re}+1]^3 > 0$. It can be concluded that the second derivative of $F(\alpha)$ is less than zero. Thus, $F(\alpha)$ is a concave function. Therefore, the maximum value of $F(\alpha)$ corresponds to the zero of the first derivative of $F(\alpha)$. Let Equation (22) be equal to 0; then, the equation for the optimal solution is

$$\alpha^* = \dfrac{\gamma_{re}+1-\sqrt{(\gamma_{re}+1)(\gamma_{sd}+1)}}{\gamma_{re}-\gamma_{sd}}$$
$$= 1+\dfrac{\gamma_{sd}+1-\sqrt{(\gamma_{re}+1)(\gamma_{sd}+1)}}{\gamma_{re}-\gamma_{sd}} \tag{24}$$

Because $\alpha^* \in (0,1)$, the value range of Equation (24) needs to be determined. When $\gamma_{re} < \gamma_{sd}$, $\gamma_{re} + 1 - \sqrt{(\gamma_{re}+1)(\gamma_{sd}+1)} < 0$ and $\gamma_{re} - \gamma_{sd} < 0$; thus, $\alpha^* > 0$. At the same time, $\gamma_{sd} + 1 - \sqrt{(\gamma_{re}+1)(\gamma_{sd}+1)} > 0$, so $\frac{\gamma_{sd} + 1 - \sqrt{(\gamma_{re}+1)(\gamma_{sd}+1)}}{\gamma_{re} - \gamma_{sd}} < 0$; therefore, $\alpha^* < 1$. Thus, when $\gamma_{re} < \gamma_{sd}$, $0 < \alpha^* < 1$. When $\gamma_{re} > \gamma_{sd}$, $\gamma_{re} - \gamma_{sd} > 0$, and thus, $\gamma_{re} + 1 - \sqrt{(\gamma_{re}+1)(\gamma_{sd}+1)} > 0$. Consequently, $\alpha^* > 0$, and also $\gamma_{sd} + 1 - \sqrt{(\gamma_{re}+1)(\gamma_{sd}+1)} < 0$, $\frac{\gamma_{sd} + 1 - \sqrt{(\gamma_{re}+1)(\gamma_{sd}+1)}}{\gamma_{re} - \gamma_{sd}} < 0$, and $\alpha^* < 1$. Therefore, $0 < \alpha^* < 1$, which corresponds to the constraints of the power allocation factor. Hence, Equation (24) is the final solution for the optimal power allocation factor.

Substituting $\gamma_{sd} = P\|\mathbf{h}_{sd}\|^2/\sigma_d^2$ and $\gamma_{re} = P\|\mathbf{h}_{re,k}^H \mathbf{G}_{\mathbf{h}_{rd,k}}\|^2/\sigma_e^2$ into Equation (24), the final equation of optimal power allocation factor α* can be obtained and is expressed as

$$\alpha^* = \frac{\frac{P}{\sigma^2}\|\mathbf{h}_{re,k}^H \mathbf{G}_{\mathbf{h}_{rd,k}}\|^2 + 1 - \sqrt{(\frac{P}{\sigma^2}\|\mathbf{h}_{re,k}^H \mathbf{G}_{\mathbf{h}_{rd,k}}\|^2 + 1)(\frac{P}{\sigma^2}\|\mathbf{h}_{sd}\|^2 + 1)}}{\frac{P}{\sigma^2}(\|\mathbf{h}_{re,k}^H \mathbf{G}_{\mathbf{h}_{rd,k}}\|^2 - \|\mathbf{h}_{sd}\|^2)}$$

(25)

Equation (25) reveals that the optimal power allocation factor α* is related to the total transmitting power constraints $P$, $\|\mathbf{h}_{sd}\|^2$ and $\|\mathbf{h}_{re,k}^H \mathbf{G}_{\mathbf{h}_{rd,k}}\|^2$. Therefore, to conduct such optimal power allocation for the satellite communication system, the instantaneous CSI of channels $\mathbf{h}_{sd}$, $\mathbf{h}_{rd,k}$ and $\mathbf{h}_{re,k}$ must be known.

Therefore, when the satellite communication system performs power allocation according to the transmitting power of Alice, α*P, and that of the relay, (1-α*)P, the minimum SOP of the system is obtained.

C. Power allocation based on statistical CSI

When the system performs optimal power allocation, the known instantaneous CSI is required. However, accurate channel estimation is also required for the satellite communication system to obtain the instantaneous CSI $\mathbf{h}_{sd}$, $\mathbf{h}_{re,k}$, and $\mathbf{h}_{rd,k}$. This makes it difficult to implement this method in actual practice; there may be circumstance in which the system is unable to attain the instantaneous CSI. Moreover, the instantaneous CSI $\mathbf{h}_{sd}$, $\mathbf{h}_{re,k}$ and $\mathbf{h}_{rd,k}$ are constantly changing; thus, the system must update the power allocation factor continuously according to changes in the instantaneous CSI. However, updating the power allocation factor increases the signaling overhead and the complexity for the system to perform power allocation.

Therefore, to compensate for the inadequacy of optimal power allocation and ensure the realizability of the method in actual practice, statistical CSI power allocation is investigated. The power allocation is performed based on the statistical CSI of the channel to determine the value of the power allocation factor. Therefore, when the statistical CSI of the channel is constant, the power allocation factor remains unchanged. Then, the SOP of the system only depends on the distribution of channels $\mathbf{h}_{sd}$, $\mathbf{h}_{re,k}$ and $\mathbf{h}_{rd,k}$. The statistical CSI-based power allocation is still intended to attain the optimal performance in terms of the SOP as much as possible; therefore, the derivation of the SOP is first conducted.

Substituting Equation (17) into Equation (7), the equation for the SOP can be expressed as

$$P_{out}(\alpha) = Pr[\gamma_{re} < \frac{(1+\alpha\gamma_{sd})(2^{R_s}-1)}{(1+\alpha\gamma_{sd}-2^{R_s})(1-\alpha)}, \gamma_{sd} > \frac{2^{R_s}-1}{\alpha}]$$
$$+ Pr[\gamma_{re} > \frac{(1+\alpha\gamma_{sd})(2^{R_s}-1)}{(1+\alpha\gamma_{sd}-2^{R_s})(1-\alpha)}, \gamma_{sd} < \frac{2^{R_s}-1}{\alpha}]$$

(26)

Expressing the above in integral form and simplifying, the SOP can be rewritten as

$$P_{out}(\alpha) = 1 + F_{\gamma_{sd}}(V_{sd}) - 2\int_0^{V_{sd}} p_{\gamma_{sd}}(\gamma_{sd}) F_{\gamma_{re}}(V_{re}) \mathrm{d}\gamma_{sd}$$

(27)

in which $p_{\gamma_{sd}}(\gamma_{sd})$ represents the probability density function (PDF) of $\gamma_{sd}$, $F_{\gamma_{sd}}(\gamma_{sd})$ represents the CDF of the random variable $\gamma_{sd}$, $p_{\gamma_{re}}(\gamma_{re})$ represents the PDF of $\gamma_{re}$, $F_{\gamma_{re}}(\gamma_{re})$ represents the CDF of $\gamma_{re}$, $V_{re}$ and $V_{sd}$ represent the integral threshold, and $V_{sd} = \dfrac{2^{R_s}-1}{\alpha}$.

It can be observed from Equation (27) that the expression for the SOP is related to the PDFs and CDFs of the random variables $\gamma_{sd}$ and $\gamma_{re}$. Thus, next, the PDFs and CDFs of $\gamma_{sd}$ and $\gamma_{re}$ are derived.

Satellite-ground channel $\mathbf{h}_{sd}$ is a Rician fading channel with very large Rician factor. Therefore, the random variable $\|\mathbf{h}_{sd}\|^2$ obeys the non-central chi-square distribution that has number of degrees of freedom equal to $2N_r$. Additionally, because $\gamma_{sd} = P\|\mathbf{h}_{sd}\|^2/\sigma_d^2$, through derivation, the PDF and CDF of $\gamma_{sd}$ are

$$p_{\gamma_{sd}}(\gamma_{sd}) = \begin{cases} \dfrac{1}{2\sigma_{sd}^2}\left(\dfrac{\gamma_{sd}}{Ps_{sd}^2\sigma_d^2}\right)^{\frac{N_r-1}{2}} \mathrm{e}^{-\frac{Ps_{sd}^2+\gamma_{sd}\sigma_d^2}{2P\sigma_{sd}^2}} I_{N_r-1}\left(\dfrac{s_{sd}}{\sigma_{sd}^2}\sqrt{\dfrac{\gamma_{sd}\sigma_d^2}{P}}\right), & \gamma_{sd} > 0 \\ 0, & \gamma_{sd} \leq 0 \end{cases} \quad (28)$$

$$F_{\gamma_{sd}}(\gamma_{sd}) = \begin{cases} 1 - Q_{N_r}\left(\dfrac{s_{sd}}{\sigma_{sd}}, \sqrt{\dfrac{\gamma_{sd}\sigma_d^2}{P\sigma_{sd}^2}}\right), & \gamma_{sd} > 0 \\ 0, & \gamma_{sd} \leq 0 \end{cases} \quad (29)$$

in which $s_{sd}^2 = \dfrac{K_{sd}E\left[\|\mathbf{h}_{sd}\|^2\right]}{(K_{sd}+1)}$ and $\sigma_{sd}^2 = \dfrac{E\left[\|\mathbf{h}_{sd}\|^2\right]}{2N_r(K_{sd}+1)}$. $K_{sd}$ is the Rician factor of the channel $\mathbf{h}_{sd}$; since the channel is a Rician fading channel with very large Rician factor, the value of $K_{sd}$ is relatively large.

(1) When $\mathbf{h}_{rd,k}$ and $\mathbf{h}_{re,k}$ are independent Rayleigh channels with same distribution, $\|\mathbf{h}_{re,k}^H \mathbf{G}_{\mathbf{h}_{rd,k}}\|^2$ obeys the $\chi^2$ distribution that has $2N_r - 2$ degrees of freedom. Additionally, because $\gamma_{re} = P\|\mathbf{h}_{re,k}^H \mathbf{G}_{\mathbf{h}_{rd,k}}\|^2/\sigma_e^2$, the CDF of $\gamma_{re}$ is the equation for the SOP,

$$F_{\gamma_{re}}(\gamma_{re}) = \begin{cases} 1 - \mathrm{e}^{-\frac{\gamma_{re}\sigma_e^2}{2P\sigma_{re}^2}} \sum_{m=0}^{N_r-2} \dfrac{1}{m!}\left(\dfrac{\gamma_{re}\sigma_e^2}{2P\sigma_{re}^2}\right)^m, & \gamma_{re} > 0 \\ 0, & \gamma_{re} \leq 0 \end{cases} \quad (30)$$

Substituting Equations (28), (29), and (30) into Equation (27), the expression for the SOP under the aforementioned conditions is

$$P_{out}(\alpha) = P_{out1} + 2P_{out2} \quad (31)$$

where $\sigma_{re}^2 = \dfrac{E\left[\|\mathbf{h}_{re}\|^2\right]}{2N_r}$;

$$P_{out1} = Q_{N_r}\left(\dfrac{s_{sd}}{\sigma_{sd}}, \sqrt{\dfrac{V_{sd}\sigma_d^2}{P\sigma_{sd}^2}}\right);$$

$$P_{out2} = \int_0^{V_{sd}} \dfrac{1}{2\sigma_{sd}^2}\left(\dfrac{\gamma_{sd}}{Ps_{sd}^2\sigma_d^2}\right)^{\frac{N_r-1}{2}} \mathrm{e}^{-\frac{Ps_{sd}^2+\gamma_{sd}\sigma_d^2}{2P\sigma_{sd}^2}} I_{N_r-1}\left(\dfrac{s_{sd}}{\sigma_{sd}^2}\sqrt{\dfrac{\gamma_{sd}\sigma_d^2}{P}}\right) \mathrm{e}^{-\frac{V_{re}\sigma_e^2}{2P\sigma_{re}^2}} \sum_{m=0}^{N_r-2} \dfrac{1}{m!}\left(\dfrac{V_{re}\sigma_e^2}{2P\sigma_{re}^2}\right)^m \mathrm{d}\gamma_{sd}$$

(2) When $\mathbf{h}_{rd,k}$ and $\mathbf{h}_{re,k}$ are independent Rayleigh channels with same distribution, $\|\mathbf{h}_{re,k}^H \mathbf{G}_{\mathbf{h}_{rd,k}}\|^2$ obeys the non-central $\chi^2$ distribution with $2N_r - 2$ degrees of freedom. Additionally, because $\gamma_{re} = P\|\mathbf{h}_{re,k}^H \mathbf{G}_{\mathbf{h}_{rd,k}}\|^2/\sigma_e^2$, the

CDF of $\gamma_{re}$ is the expression for the SOP:

$$F_{\gamma_{re}}(\gamma_{re}) = \begin{cases} 1 - Q_{N_r-1}(\frac{s}{\sigma_{re}}, \sqrt{\frac{\gamma_{re}\sigma_e^2}{P\sigma_{re}^2}}), & \gamma_{re} > 0 \\ 0, & \gamma_{re} \leq 0 \end{cases} \quad (32)$$

where $s_{re}^2 = \frac{K_{re,k}E\left[\|\mathbf{h}_{re,k}\|^2\right]}{(K_{re,k}+1)}$, $\sigma_{re}^2 = \frac{E\left[\|\mathbf{h}_{re,k}\|^2\right]}{2N_r(K_{re,k}+1)}$,

and $K_{re,k}$ is the Rician factor of channel $\mathbf{h}_{re,k}$.

Substituting Equations (28), (29), and (32) into Equation (27), the expression for the SOP under the aforementioned conditions is

$$P_{out}(\alpha) = P_{out1} + 2P_{out3} \quad (33)$$

where

$$P_{out3} = \int_0^{V_{ud}} \frac{1}{2\sigma_{sd}^2} \left(\frac{\gamma_{sd}}{Ps_{sd}^2\sigma_d^2}\right)^{\frac{N_r-1}{2}} e^{-\frac{Ps_{sd}^2+\gamma_{sd}\sigma_d^2}{2P\sigma_{sd}^2}} I_{N_r-1}\left(\frac{s_{sd}}{\sigma_{sd}^2}\sqrt{\frac{\gamma_{sd}\sigma_d^2}{P}}\right) Q_{N_r-1}\left(\frac{s_{sd}}{\sigma_{re}}, \sqrt{\frac{\gamma_{re}\sigma_e^2}{P\sigma_{re}^2}}\right) d\gamma_{sd}.$$

It can be observed from Equations (31) and (33) that regardless of whether the channel between the relay and the ground receiving end, $\mathbf{h}_{rd,k}$ and $\mathbf{h}_{re,k}$, is a Rayleigh channel or a Rician channel, the SOP expression is not of a closed form, which makes it difficult to solve for the optimal power distribution factor. However, because the channel between the satellite and the ground receiving end is a Rician fading channel with very lager Rician factor in practice, the Rician distribution approximates the Gaussian distribution when the Rician factor is very large; that is, a Rician channel is approximately a Gaussian channel. Therefore, to obtain a closed-form SOP expression for calculation convenience, we directly assume Gaussian channels to approximate Rician fading channels $\mathbf{h}_{sd}$ and $\mathbf{h}_{se}$ in the process of the SOP expression derivation.

Assume that the gains of the $\mathbf{h}_{sd}$ and $\mathbf{h}_{se}$ channels are equal to $A$, $A = E[\|\mathbf{h}_{sd}\|^2]$; then, the following expression for the instantaneous secrecy capacity can be obtained by substituting $\|\mathbf{h}_{sd}\|^2 = A$ into Equation (17) and simplifying:

$$C_{s,k}(\alpha) = \log_2\left(\frac{1 + \frac{\alpha PA}{\sigma_d^2}}{1 + \frac{\alpha PA}{(1-\alpha)P\|\mathbf{h}_{re,k}^H \mathbf{G}_{\mathbf{h}_{rd,k}}\|^2 + \sigma_e^2}}\right) \quad (34)$$

Substituting Equation (34) into Equation (7) and simplifying, the expression for the SOP is

$$P_{out}(\alpha) = \begin{cases} F_{\gamma_{re}}\left(\frac{(\alpha P'A+1)(2^{R_s}-1)}{(1+\alpha P'A-2^{R_s})(1-\alpha)}\right), & \alpha > \frac{2^{R_s}-1}{P'A} \\ 1, & \alpha = \frac{2^{R_s}-1}{P'A} \\ 1 - F_{\gamma_{re}}\left(\frac{(\alpha P'A+1)(2^{R_s}-1)}{(1+\alpha P'A-2^{R_s})(1-\alpha)}\right), & \alpha < \frac{2^{R_s}-1}{P'A} \end{cases} \quad (35)$$

where $P' = \frac{P}{\sigma_d^2}$.

Substituting Equation (30) into Equation (35), the following approximate closed-form expression for the SOP of $\mathbf{h}_{rd,k}$ and $\mathbf{h}_{re,k}$, when they are independent and identically distributed Rayleigh channels, can be obtained:

$$P_{out}^{Re}(\alpha) = \begin{cases} 1 - e^{-\frac{(\alpha P'A+1)(2^{R_s}-1)}{2P'(1-\alpha)(1+\alpha P'A-2^{R_s})\sigma_{re}^2}} \sum_{m=0}^{N_r-2} \frac{1}{m!}\left(\frac{(\alpha P'A+1)(2^{R_s}-1)}{2P'(1-\alpha)(1+\alpha P'A-2^{R_s})\sigma_{re}^2}\right)^m, & \alpha > \frac{2^{R_s}-1}{P'A} \\ 1, & \alpha = \frac{2^{R_s}-1}{P'A} \\ e^{-\frac{(\alpha P'A+1)(2^{R_s}-1)}{2P'(1-\alpha)(1+\alpha P'A-2^{R_s})\sigma_{re}^2}} \sum_{m=0}^{N_r-2} \frac{1}{m!}\left(\frac{(\alpha P'A+1)(2^{R_s}-1)}{2P'(1-\alpha)(1+\alpha P'A-2^{R_s})\sigma_{re}^2}\right)^m, & \alpha < \frac{2^{R_s}-1}{P'A} \end{cases} \quad (36)$$

Substituting Equation (32) into Equation (35), the following approximate closed-form expression for the SOP of $\mathbf{h}_{rd,k}$ and $\mathbf{h}_{re,k}$, when they are independent and identically distributed Rician channels, can be obtained:

$$P_{out}^{Ri}(\alpha) = \begin{cases} 1 - Q_{N_r-1}\left(\frac{s_{re}}{\sigma_{re}}, \sqrt{\frac{(\alpha P'A+1)(2^{R_s}-1)}{(1+\alpha P'A-2^{R_s})(1-\alpha)P'\sigma_{re}^2}}\right), & \alpha > \frac{2^{R_s}-1}{P'A} \\ 1, & \alpha = \frac{2^{R_s}-1}{P'A} \\ Q_{N_r-1}\left(\frac{s_{re}}{\sigma_{re}}, \sqrt{\frac{(\alpha P'A+1)(2^{R_s}-1)}{(1+\alpha P'A-2^{R_s})(1-\alpha)P'\sigma_{re}^2}}\right), & \alpha < \frac{2^{R_s}-1}{P'A} \end{cases}$$

(37)

Based on Equations (36) and (37), directly approximating the Rician fading channels with very large Rician factor $\mathbf{h}_{sd}$ and $\mathbf{h}_{se}$ as Gaussian channels in the derivation, a closed-form expression for the SOP can be obtained. Thus, as long as the statistical CSI of channels $\mathbf{h}_{sd}$ and $\mathbf{h}_{re,k}$ is known, the SOP at location of Bob can be directly and rapidly calculated. Moreover, because the range of the power allocation factor $\alpha$ value, $\alpha \in (0,1)$, is quite small, the traversal search method can be employed. It is thus possible to traverse the power allocation factor $\alpha$ according to a specified step size, calculate the SOP, and select the $\alpha$ value resulting in the minimum SOP as the final power allocation factor; the specific algorithm is detailed in Table 1.

Table 1 CSI-based power allocation factor optimization algorithm flow

| **Algorithm 1 CSI-based power allocation factor optimization algorithm flow** |
|---|
| **Input:** |
|     Traversing step size of $\alpha$: $\Delta\alpha$ |
|     Total emission power: $P$ |
|     Bob and Eve receiving end noise power: $\sigma^2$ |
|     $\mathbf{h}_{sd}$ and $\mathbf{h}_{re}$ statistical CSI: $E\left[\|\mathbf{h}_{sd}\|^2\right]$, $E\left[\|\mathbf{h}_{re}\|^2\right]$ |
|     Rician factor of $\mathbf{h}_{re}$ (this parameter does not exist when $\mathbf{h}_{re}$ is Rayleigh channel): $K_{re}$ |
| **Output:** Power allocation factor $\alpha$ |
| 1. Calculate all traversal values of power allocation factor $\alpha$. |
|     Quantify the possible values of $\alpha$ in the interval $(0,1)$, and store as one-dimensional vector $\mathbf{B}$ |
| $$\mathbf{B}(i) = \Delta a \cdot i, \quad i = 1,..., \frac{1}{\Delta a} - 1$$ |
| 2. Travers power allocation factor to calculate SOP. |
|     According to the closed form expression of the power allocation factor, calculate the SOP for power allocation factor being every element in $\mathbf{B}$, and store as one-dimensional vector $\mathbf{P}_{out}$. |
| Rayleigh channel $\mathbf{h}_{re}$: |
| $$\mathbf{P}_{out}(i) = P_{out}^{Re}(\mathbf{B}(i))$$ |
| Rician channel $\mathbf{h}_{re}$: |
| $$\mathbf{P}_{out}(i) = P_{out}^{Ri}(\mathbf{B}(i))$$ |
| 3. Obtain final power allocation factor $\alpha$. |
| $$i_{op} = \underset{i=1,...,\frac{1}{\Delta a}-2}{\arg\min} \mathbf{P}_{out}(i)$$ |
| $$\alpha = \mathbf{B}(i_{op})$$ |

## 5 Numerical Results and Analysis

A. Performance analysis of SOP for relay selection

An analysis of the relay antenna number $N_r$ is shown in Figure 2, which shows the variation curve of the SOP of the two relay selections with respect to the satellite and relay sum transmission power constraint $P$ when $N_r = 2, 4, 8$. Assume that the minimum rate of secure transmission of the system $R_s$ is 2 bit/s/Hz, $N_s = N_r$, the noise power at the receiving end is normalized to 1, $P = P/\sigma^2$, and $P$ is 5-15 dB. To conform to the actual situation of the satellite-ground channel, the channels between the satellite emitting end and ground receiving end, $\mathbf{h}_{sd}$ and $\mathbf{h}_{se}$, are Rician fading channels with Rician factor $K_{sd} = 10$. The channels between the relay and ground receiving ends, $\mathbf{h}_{re}$ and $\mathbf{h}_{rd}$, are Rayleigh channels.

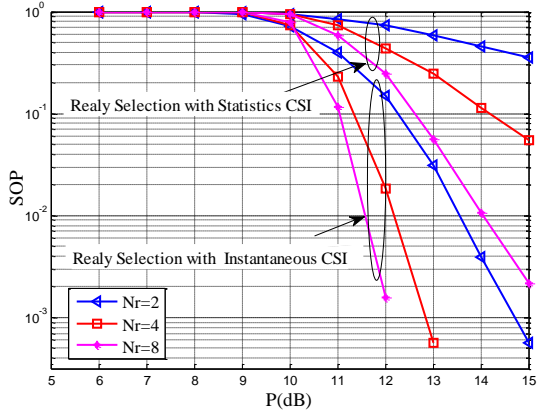

Figure 2 SOP variation curve with respect to the total power *P* in each relay plan when $N_r = 2, 4, 8$

It can be observed from Figure 2 that when all other conditions are fixed, the SOP performance of the selected relay in the condition of known instantaneous CSI is far better than that in the condition of known statistical CSI. This result indicates that the relay selection achieved the enhancement of the SOP performance. In addition, when all other conditions are fixed, the higher the number $N_r$, the better the SOP performance of the two relay selections. For example, for the relay selection under the condition of known instantaneous CSI in Figure 2, when $P = 12\text{dB}$ and $N_r=2$, the order of magnitude of the SOP is greater than $10^{-1}$; it is $10^{-1}$ when $N_r=8$, and it is $10^{-2}$ when $N_r=8$. Additionally, in Figure 2, the SOP of the relay selection based on the instantaneous CSI when $N_r=2$ is far less than that based on the statistical CSI when $N_r=8$. This result once again indicates that the SOP performance of the selected relay under the condition of known instantaneous CSI is far better than that under the condition of known statistical CSI.

Figure 3 shows the impact of the minimum rate of secure transmission of the system $R_s$ on the SOP performance of the relay selection; the figure shows the variation curve of the SOP of each relay selection with respect to the total power constraint *P* when $R_s = 1, 1.5, 2$. The condition of $N_s = N_r = 4$ is assumed here, and the other simulation conditions are the same as in Figure 2.

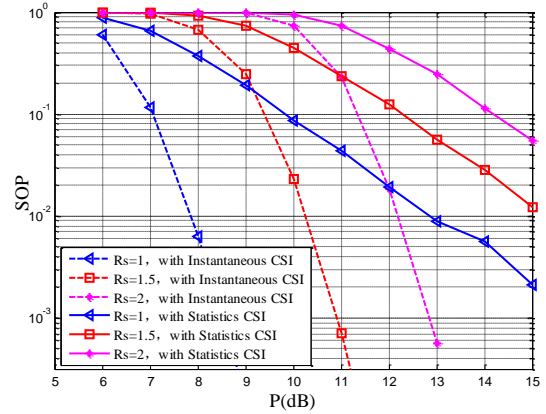

Figure 3 Variation curve of the SOP of each relay selection with respect to the total power constraint P when $R_s = 1, 1.5, 2$

It can also be observed from Figure 3 that the SOP performance of the selected relay with known instantaneous CSI is far better than that with known statistical CSI. In addition, as $R_s$ decreases, the SOP performances of both relay selections improve, but the improvement in the relay selection with known instantaneous CSI is superior to that with known statistical CSI. For instance, in Figure 3, when $P = 11\text{dB}$ and $R_s$ decreased from 2 to 1.5 bit/s/Hz, the order of magnitude of the SOP of the relay selection with known instantaneous CSI decreased from $10^{-1}$ to $10^{-3}$, whereas that of relay selection with known statistical CSI did not change.

B. Relevant statistical analysis based on power allocation using statistical CSI

The SOP of the statistical CSI-based power allocation is affected by the traversal step size used to determine the power allocation factor and the value of the Rician factor of the actual satellite-to-ground receiving end channel. Therefore, the impact of the two on the SOP of the statistical CSI-based power allocation is analyzed via a simulation.

Figure 4 shows the analysis of the influence of the Rician factor $K_{sd}$ of the actual satellite-ground channel $\mathbf{h}_{sd}$ on the SOP performance of the statistical CSI-based power allocation; it shows the variation curves of the SOP of each power allocation method with respect to the Rician factor $K_{sd}$ of the channel $\mathbf{h}_{sd}$. Assume that the system's minimum secure transmission rate $R_s$ is 1 bit/s/Hz, $N_d = N_r = 4$, the receiving end noise power is normalized to 1, $P = P/\sigma^2$, $P = 10$ dB, and when the channel between the relay and the ground receiving end is a Rician channel, the Rician factor $K_{re}$ is 1. Then, $E[\|\mathbf{h}_{sd}\|^2] = E[\|\mathbf{h}_{re}\|^2] = 1$. It can be observed from Figure 4 that regardless of whether the channel between the relay and the ground receiving end is a Rician or Rayleigh channel, for higher $K_{sd}$, the SOP performance of the statistical CSI-based method is closer to the optimal SOP performance. When $K_{sd}$ is relatively small, the SOP of the statistical CSI-based power allocation method is quite different from the optimal one and even worse than the SOP performance of the uniform allocation method. For example, in Figure 4, when the channel between the relay and ground receiving end is a Rayleigh channel and $K_{sd} \leq 2$, the SOP of the statistical CSI-based power allocation method is greater than the SOP of the uniform allocation method, whereas when $K_{sd} \geq 12$, the SOP of the statistical CSI-based power allocation method is almost equal to that of optimal power allocation. This result occurred because we treated channel $\mathbf{h}_{sd}$ as a Gaussian channel rather than the actual Rician fading channel with large Rician factor for approximate optimization in the statistical CSI-based power allocation approach. Therefore, the closer the actual channel $\mathbf{h}_{sd}$ is to a Gaussian channel, the closer the statistical CSI based power allocation is to the optimal solution; the larger the Rician factor of a Rician channel, the closer the channel is to a Gaussian channel. In reality, the Rician factor of the channel between the satellite and ground receiving end is quite large; therefore, the statistical CSI-based power allocation method is feasible.

In addition, when the Rician factor $K_{sd}$ of the satellite-ground channel $\mathbf{h}_{sd}$ is relatively small, the SOP of the statistical CSI-based power allocation method is closer to the SOP of the optimal power allocation when the channel between the relay and the ground receiving end is a Rician channel, compared to when the channel is a Rayleigh channel. Additionally, when the other conditions are fixed, the SOP performance of the relay-to-ground receiving end channel is better when it is a Rician channel than when it is a Rayleigh channel.

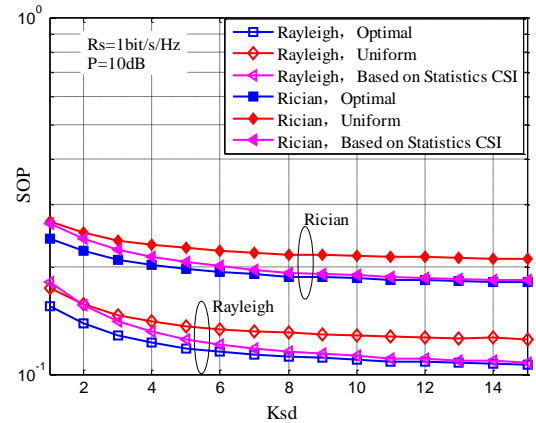

Figure 4 Variation curve of the SOP of the power allocation method with respect to the Rician factor $K_{sd}$

of channel $\mathbf{h}_{sd}$

Figure 5 analyzes the effect of the traversal step size on the SOP performance of the statistical CSI-based power allocation method; it shows the statistical CSI-based power allocation scheme curve whose SOP varies with the total power constraint P in the cases in which the relay-ground receiving end channel is a Rician and a Rayleigh channel ($\Delta a = 0.001, 0.005, 0.01, 0.05, 0.1$). Assume that the minimum rate of the system secure transmission $R_s$ is 1 bit/s/Hz, $N_s = N_r = 4$, the noise power of the receiving end is normalized to 1, $P = P/\sigma^2$, $P$ is in the range of 5 to 15 dB, and the satellite-legitimate user channel and satellite-ground receiving end channel are both Rician channels with Rician factors $K_{sd}$ of 10. When the relay-ground receiving end channel is Rician, its Rician factor $K_{re,k}$ is 1. Then, $E[\|\mathbf{h}_{sd}\|^2] = E[\|\mathbf{h}_{re}\|^2] = 1$.

Figure 5 reveals that regardless of whether the relay-ground receiving end channel is Rician or Rayleigh, for $\Delta a = 0.001, 0.005, 0.01, 0.05, 0.1$, the performance gap in terms of the SOP is small for the statistical CSI-based power allocation method. For $\Delta a = 0.1$, i.e., high signal-noise ratio, the SOP is slightly greater than the other four, and the other four curves almost coincide. From this result, it can be noted that the traversal-step-length precision of power allocation factor reaching 0.05 can approximately obtain the optimal SOP. Since the value range of the power allocation factor α is $\alpha \in (0,1)$, with a traversal step size of 0.05, only 19 traversals must be conducted. Meanwhile, the SOP has a closed-form expression that can be directly calculated. Therefore, the computational complexity of the statistical CSI-based power allocation method is achievable by the system.

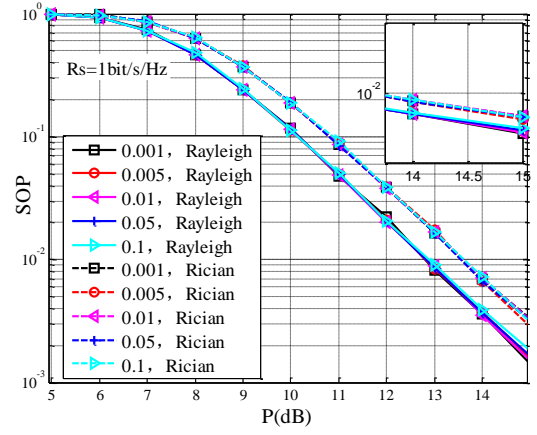

Figure 5 Variation curve of the statistical CSI-based power allocation SOP with respect to the total power constraint $P$

C. Comparison and analysis of each power allocation scheme in terms of the SOP performance

To perform a better comparison of the secrecy performance for each power allocation, the SOP of each power allocation is simulated and analyzed. At the same time, the influence of the number of relay antennas $N_r$ and the minimum rate of system secure transmission $R_s$ on the performance of the SOP are analyzed via simulations.

Figure 6 analyzes the effect of the number of relay antennas $N_r$ on the SOP for each power allocation; it shows the variation curve of the SOP of each power allocation method with respect to the total power constraint $P$ when $N_r = 2, 4, 8$. Assume that the minimum rate of secure transmission $R_s$ is 1 bit/s/Hz, $N_s = N_r$, the noise power of the receiving end is normalized to 1, $P = P/\sigma^2$, and the total transmission power constraint P is in the range from 5 to 20 db. To correspond with the actual situation of the satellite-ground channel, the channel of satellite-transmitting end to the ground-receiving end is a Rician fading channel with Rician factor $K_{sd} = 5$. Figure 6 a) shows the situation in which the relay-to-ground receiving end channel is a Rician channel, whereas Figure 6 b) shows the situation in which the channel is a Rayleigh channel with Rayleigh factor $K_{re,k} = 1$.

It can be noted from Figure 6 that regardless of whether the relay-to-ground receiving end channel is a Rician or Rayleigh channel, when the other conditions are fixed, the performance optimal power allocation SOP is the best. The performance of statistical CSI-based SOP is second, but it is very close to that of the optimal power allocation SOP. In addition, the SOP performance of the two power allocations is far superior to that of uniform power allocation, which indicates that the two proposed power allocation methods enhanced the performance in terms of the system SOP. When the other conditions are fixed, greater $N_r$ corresponds to smaller SOP values for each power allocation method and thus better performance. For example, in Figure 6 a), at $P = 15\text{dB}$ and $N_r = 2$, the order of magnitude of the SOP is $10^{-1}$. When $N_r = 4$, the order of magnitude is $10^{-2}$, and at $N_r = 8$, the order of magnitude is less than or equal to $10^{-5}$.

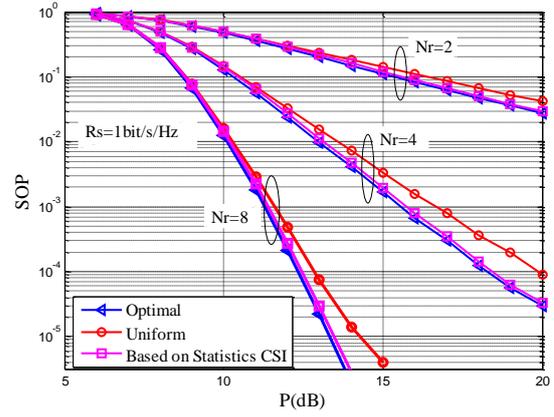

a) Situation of relay to the ground when the receiving end channel is a Rayleigh channel

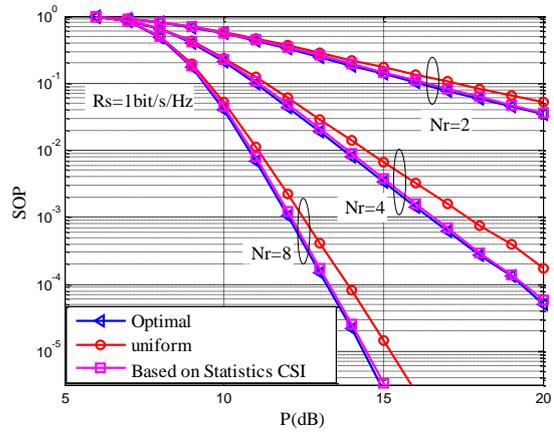

b) Situation of relay to the ground when the receiving end channel is a Rician channel

Figure 6 Variation curve of the SOP of each power allocation method with respect to the total power constraint $P$ when $N_r = 2, 4, 8$

Figure 7 analyzes the effect of the minimum security transmission rate $R_s$ on the performance of the power allocation SOP and shows the variation curve of power allocation of each method with respect to the total power constraint $P$. It is assumed that $N_s = N_r = 4$, and the other conditions are the same as those in Figure 6.

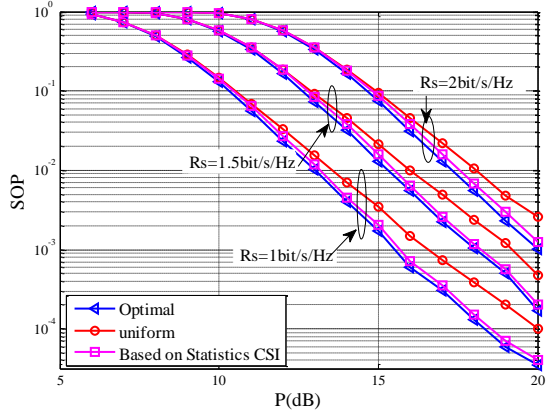

a) Situation of relay to the ground when the receiving end channel is a Rayleigh channel

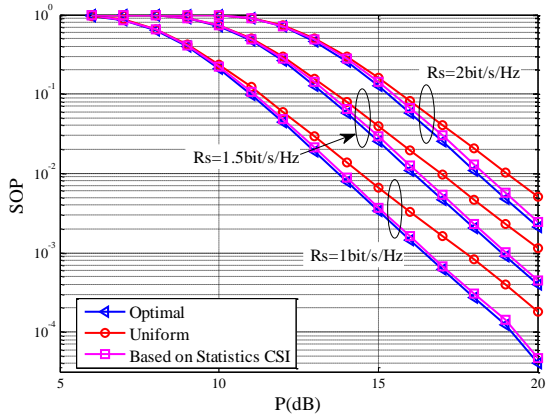

b) Situation of relay to the ground when the receiving end channel is a Rician channel

Figure 7 Variation curve of the SOP for each power allocation method with respect to the total power constraint $P$ when $R_s = 1, 1.5, 2$

It can be noted from the curves in Figure 7 that regardless of whether the relay-to-ground receiving end channel is a Rician or Rayleigh channel, the smaller the value of $R_s$, the smaller the SOP. For example, in Figure 7 b), when $P$=15 dB and $R_s = 1$, the order of magnitude for SOP is $10^{-2}$. At $R_s = 1.5$, the SOP order of magnitude is $10^{-1}$, and at $R_s = 2$, the SOP order of magnitude is greater than $10^{-1}$. Taking the derivative of Equation (7) yields

$$P_{out,k}(\alpha) = Pr[C_{s,k}(\alpha) < R_1]$$
$$= Pr[C_{s,k}(\alpha) < R_2] + Pr[R_2 \leq C_{s,k}(\alpha) < R_1]$$
$$> Pr[C_{s,k}(\alpha) < R_2], \ (R_1 > R_2)$$

The above yields the same result as the simulation.

**6 Conclusions**

This paper studied secure physical layer communication in satellite communication via similar channels. A satellite-ground physical layer security communication model based on cooperative interference relay is proposed, and its feasibility is theoretically analyzed. The optimization of the relay selection standard is conducted on the basis of the model for the purpose of improving the performance of the system SOP. This approach specifically includes the relay selection criteria for conditions of known instantaneous CSI and known statistical CSI. Additionally, based on the model, when the total transmitting power of the satellite and relay is fixed, the power allocation between the satellite and the relay is optimized, which gives the optimal power allocation that minimizes the SOP. Additionally, a statistical CSI-based power allocation that considers realizability was proposed. The simulation results indicate that the performance of the SOP obtained with the known instantaneous CSI relay selection method exceeds that based on the known statistical CSI. Nevertheless, power allocation based on the statistical CSI is achievable despite its complexity. In addition, the performance of the optimal power allocation SOP is superior to that of statistical CSI-based power allocation, and the performance of statistical CSI-based power allocation is better than that of uniform power allocation.

References

[1] Ammari M L, Fortier P. Physical Layer Security of Multiple Input multiple output systems with transmit beamforming in Rayleigh fading[J]. Iet Communications,